\documentclass[aps,prl,reprint,twocolumn,amsmath,amssymb,citeautoscript,longbibliography]{revtex4-1}
\usepackage{graphicx}
\usepackage{dcolumn}
\usepackage{bm}
\usepackage{latexsym,epsfig}
\usepackage{graphicx}
\usepackage{verbatim}
\usepackage{comment}
\usepackage{amsmath}
\usepackage{amssymb}
\usepackage{physics}
\usepackage{stmaryrd}
\usepackage{color}
\usepackage{epstopdf}
\usepackage{grffile}
\usepackage{lipsum}
\usepackage{enumitem}
\usepackage{ulem}

\usepackage[dvipsnames]{xcolor}
\DeclareGraphicsExtensions{.eps}

\newcommand{\beq}{\begin{equation}}
\newcommand{\eeq}{\end{equation}}
\newcommand{\bea}{\begin{eqnarray}}
\newcommand{\eea}{\end{eqnarray}}
\newcommand{\ben}{\begin{eqnarray*}}
\newcommand{\een}{\end{eqnarray*}}
\newcommand{\bfig}{\begin{figure}}
\newcommand{\efig}{\end{figure}}

\usepackage{hyperref}
\hypersetup{
    colorlinks=true,      
    urlcolor=blue,
    citecolor=blue,
    linkcolor=blue
}

\begin{document}
\title{Anomalous spectrum in a non-Hermitian quasiperiodic chain} 

\author{Soumya Ranjan Padhi $^{1,2}$, Sanchayan Banerjee $^{1,2}$, Tanay Nag $^3$ and Tapan Mishra $^{1,2}$}

\email{mishratapan@niser.ac.in}

\affiliation{$^1$ School of Physical Sciences, National Institute of Science Education and Research, Jatni 752050, India}

\affiliation{$^2$ Homi Bhabha National Institute, Training School Complex, Anushaktinagar, Mumbai 400094, India}

\affiliation{$^3$Department of Physics, BITS Pilani-Hyderabad Campus, Telangana 500078, India}

\date{\today}

\begin{abstract}
The spectra of particles in disordered lattices can either be completely extended or localized or can be intermediate which hosts both the localized and extended states separated from each other. In this work, however, we show that in the case of a one dimensional lattice with long-range hopping and non-Hermitian quasiperiodic onsite potential, the localized and extended states in the spectrum are intermixed with each other rather than well separated. 
As a result, an atypical intermediate phase appears where consecutive pairs of extended states intermittently appear in the pool of localized states.  
We also argue that such anomalous spectral intermixing can be realized in the short-range hopping limit by appropriate engineering of the onsite potential. Moreover, we obtain that the nature of the spectrum also reveals non-standard scenarios in the complex energy plane where the complex energies encircle real ones.
These findings shed light on the intricate interplay between the non-Hermiticity  and quasiperiodic disorder in the system.  

\end{abstract}

\maketitle

\paragraph*{Introduction.-} 
\label{sec:intro}
One of the most striking features of quasiperiodic lattices is the delocalization to localization transition of the single particle spectrum in low dimensional lattices in contrast to the lattices with random disorder where any infinitesimal disorder render localization of the states~\cite{Anderson_1958, Gang_of_four_1979}. The paradigmatic Aubry-Andr\'e (AA) model which describes a tight binding lattice with onsite quasiperiodic potential, is the simplest to exhibit localization transition. In this case an increase in the quasiperiodic potential strength leads to the localization of the entire extended single particle states after a critical potential strength~\cite{Aubry_1980, Paredesreview_2019, Roati2008}. However, moving away from the tight binding limit~\cite{ME_bich_sarma_2017, ME_Incomm_Opt_sarma2010, Roy_prl_2021, Pedro_2023, PRB_GME_sarma_2023,Loc_AA_non-nerest, Santos_prl_2019, Auditya_2021,Tanay_PRR_2020, Tanay_PRE_2020} or by generalizing the form of the quasiperiodic potential~\cite{ME_IP_sarma2020, ME_Shallow_1DQP_potentials, NN_TB_sarma2015, padhan_prbl_2022, Immanuel_prl_2018, Xiong_2020,subrotoreview_2017,expt_Int_ME_GAA}, instead of a sharp localization transition of the entire spectrum, an intermediate phase appears hosting both localized and extended states separated from each other by an energy dependent mobility edges (MEs).

\begin{figure*}[!t]
\centering
\includegraphics[width=2.0\columnwidth]{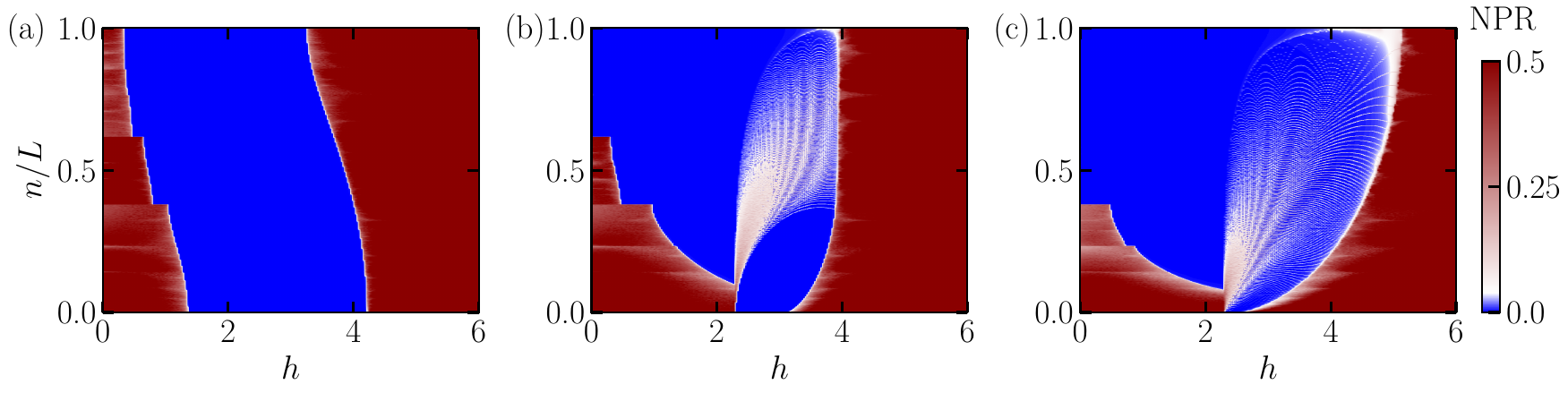}
\caption{Eigenstate index $(n/L)$ as a function of $h$ with their corresponding NPR values for power law index (a) $a=10$, (b) 1.5 and (c) 0.5 calculated in periodic boundary condition (PBC). Here,  $\beta=4181/6765$, $J=1$, $\lambda=1$, $\alpha=0.2$ and system size $L=6765$. The color-bar is appropriately set for better visibility.}
\label{fig:fig1}
\end{figure*}

More exotic features appear in the spectra of non-Hermitian systems which includes the appearance of complex eigenvalues in the spectrum, exceptional points, non-Hermitian skin effect, non-trivial spectral topology, absence of bulk-boundary correspondence, etc~\cite{Yao_prl_2018, Yuce_2020, Masatoshi_prl_2020, Li_2020, Okuma_2023, Li_2021, Lin_2023, Paolo_2023, Robert_2020, Emil_2021, Andre_2019, Nori_2019, Alvarez_2018, Masahito_prx_2018, Emil_2018, Koch_2020, Cao_2021, Xiao_2020, Jin_2019} which have been the topics of paramount interest in recent years.
In the context of quasiperiodic systems, sharp localization transition appears in the case of non-Hermitian AA models by introducing non-Hermiticity either through non-reciprocal hopping or through a complex phase factor in the quasiperiodic potential~\cite{Shuchen_2019, Longhi_prl_2019}. While for the former case the localization transition occurs as a function of the quasiperiodic potential, for the latter case, the transition occurs upon varying the complex phase, for a fixed quasiperioidic disorder strength ~\cite{Longhi_prl_2019}.  Moreover, if generalized forms of the quasiperiodic potential with complex phase factor are considered, the localization transitions are known to occur through intermediate phases hosting MEs - a situation analogous to the Hermitian cases. However, the most fundamental difference here is the localized states exhibiting complex energies due to broken parity-time $(\mathcal{PT})$ symmetry in the system. Such difference of character between the states results in loop like MEs (mobility rings) in the complex plane that distinctly separate the localized and extended states~\cite{Zhang_pra_2021, Cai_2022, Shuchen_es_2021, Zhou_prb_2023, Soutang_rent_2023, Zhou_dimerization_2022, Wu_iop_2021, Zeng_prb_2020, ShuChen_EMEs_2021, ShuChen_Exp_decay_2021, Wang_unconventional_2021, Longhi_maryland_2021, Zhou_floquet_2021, Zhou_marryland_2022, Zeng_GMEs_NHGAA_expt_2020, Li_mobrings_2024, Padhan_prbl_nh_2023, Gandhi_kitaev_2024, Peng_power_nh_2023, ShuChen_power_nh_2021, Tong_liu_2020}. Adding to these rich spectral landscape, recent studies
have also identified anomalous mobility edges~\cite{AMES_arxiv_2024} that serve
as boundaries between localized and multifractal states,
thereby enhancing the complexity of phase diagrams in
non-Hermitian quasiperiodic systems. Due to the recent advancement in the experimental front numerous such findings have been observed in various platforms such as photonic lattices~\cite{Tong_liu_2020, Longhi_prl_2019}, ultracold atoms~\cite{Zhou, BoYan, Ren_2022}, electric circuits~\cite{Zeng_GMEs_NHGAA_expt_2020, Liu_circuit_2023, Zhang2023}, periodically driven systems, etc~\cite{Weidemann2022}.

Conventionally it is understood that the spectra of quasiperiodic lattice models (whether Hermitian or non-Hermitian) constitute well separated localized and extended states even if the system is in the intermediate phases.
However, in this work we show that in the case of a non-Hermitian quasiperiodic lattice a non-trivial spectrum emerges featuring intermixed extended and localized states. By considering a one-dimensional lattice with long-range hopping and generalized Aubry Andr\'e potential with complex phase (also known as the non-Hermitian GAA or the nHGAA model) we reveal that in the spectrum each extended state are well separated from each other by a bunch of localized states giving rise to an almost regular pattern in the appearance of the extended states. Such a pattern mimics a comb like feature in the spectrum and we call this effect the non-Hermitian comb effect (NHCE). This anomalous and conventionally different spectral behavior results in more exotic and atypical distribution of eigenvalues in the complex plane where the real energies are encircled by loops formed by the complex energies. We also reveal that much richer NHCE can be obtained in the limit of short-range hopping by suitable engineering of the quasiperiodic potential. In the following we discuss these anomalous spectral behaviour in detail.


\paragraph*{Model.-}
\label{sec:model}

The Hamiltonian for the long-range nHGAA model is described by,
\begin{align}
	\mathcal{H} &= -\sum_{j,k\neq j} \big{(}\frac{J}{r_{jk}^a} \hat{c}^\dagger_j \hat{c}_k + H.c.\big{)} \nonumber \\ &+ \lambda \sum_{j} \frac{cos(2\pi\beta j + \phi)}{1 - \alpha cos(2\pi\beta j + \phi)} \hat{n}_j
 	\label{eq:Eq1}
\end{align}
Here, the creation (annihilation) operator of spinless fermions is represented by the $\hat{c}_j^\dagger (\hat{c}_j)$ at site $j$. The number operator at site $j$ is given by $\hat{n}_j$. The strength of the hopping amplitude between sites $j$ and $k$ is $J/r^a_{jk}$, where $r_{jk}$ is the distance between two sites and $a$ is power-law index that controls how the hopping strength decreases with distance. The second term in the Hamiltonian is the nHGAA potential with quasiperiodic potential strength $\lambda$. To ensure quasiperiodicity, here $\beta$ is defined as the ratio of two consecutive numbers from the Fibonacci sequence.
The non-Hermiticity in the potential is introduced by defining $\phi=\theta + ih$ and we choose $\theta=0$ for which the Hamiltonian in Eq.~\ref{eq:Eq1} is $\mathcal{PT}$ symmetric. For our studies we fix $J=1$ which sets the unit of the energy for the system. Also we consider $\lambda=1$ which is of the order of $J$ and the  system size of $L=6765$ unless otherwise mentioned. All the calculations are performed under periodic boundary condition (PBC).

\paragraph*{Results.-}
We start the discussion from the limit of $a\gg 1$ for which the model shown in Eq.~\ref{eq:Eq1} can effectively be assumed as a short-range generalized AA model with complex phase factor. It has already been shown that the tight-binding generalized AA model with complex phase exhibits a counter-intuitive phenomenon where an initially extended spectrum first localizes and then a delocalization of the entire spectra occurs as a function of $h$ for a range of values of $\alpha$ when the strength of disorder is fixed~\cite{Padhan_prbl_nh_2023}. These delocalization to localization  and localization to delocalization transitions occurs through two intermediate phases with MEs. However, for  larger values of $\alpha$, a transition from an intermediate phase to extended phase occurs. Before moving further we first establish this connection between the limit of $a\gg 1$ and the tight-binding nHGAA limit. 

To this end we compute the normalized participation ratio, $\text{NPR}_m=1 / (L \times \text{IPR}_m)$, where IPR$_m$ is the inverse participation ratio defined as $\text{IPR}_m=\sum_{j=1}^L |\psi_j^m|^4$ and $\psi_j^m$ are the probability amplitude of the individual eigenstates $m$ at site $j$~\cite{Paredesreview_2019}. This definition suggests that for the extended (localized) states, the NPR (IPR) becomes finite (zero). In Fig.~\ref{fig:fig1}(a) we plot the NPR for all the states as a function of $h$ for $a=10$ and $\alpha=0.2$ while fixing $\lambda=1$. It can be clearly seen that the entire spectrum is initially extended (NPR finite) and then an intermediate phase appears between $0.325 \lesssim h \lesssim 1.35$ where both extended (finite NPR) and localized (vanishing NPR) coexists. Further increases in $h$ results in a completely localized spectrum up to $h \approx 3.25$. After $h \gtrsim 3.25$, another intermediate phase appears between $3.25 \lesssim h \lesssim 4.25$ and eventually the entire spectrum turns extended. The MEs in the intermediate phases can be seen as the edges between the red and blue regions. This complete delocalization of the already localized spectrum in this case is similar to the case of the tight binding limit of the model shown in Eq.~\ref{eq:Eq1} and the reason behind this unusual behavior is the specific form of the potential whose strength first increases and then decreases, eventually approaching a constant values as a function of $h$~\cite{Padhan_prbl_nh_2023}. However, with increase in  the effect of the long-range hopping or by reducing the value of $a$, a surprisingly different scenario emerges which results in a different type of intermediate phase. In particular, we obtain that for a range of values of $h$, an intermediate phase appears in which a portion of the spectrum exhibits an intermixed pattern of the localized and extended states. This feature can be seen as the appearance of the states with finite NPR in Fig.~\ref{fig:fig1}(b) for $2.3 \lesssim h \lesssim 3.9$ which is plotted for $a=1.5$ and $\alpha=0.2$. 
Further decreasing the values of $a$ results in the enlarged comb region which can be seen from Fig.~\ref{fig:fig1}(c) where  $a=0.5$ is considered.

\begin{figure}[!t]
\centering
\includegraphics[width=1.0\columnwidth]{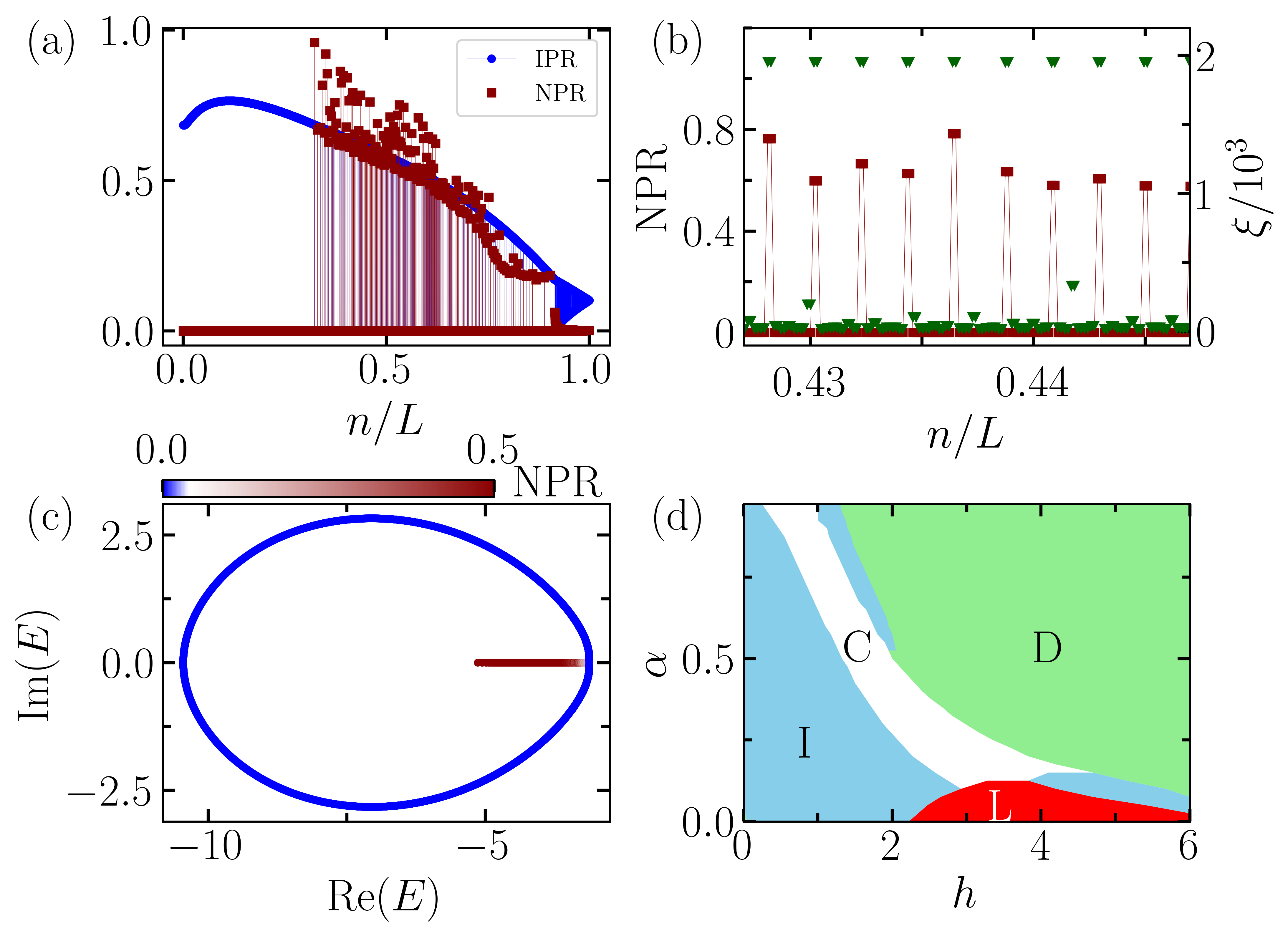}
\caption{(a) The IPR (blue circles) and NPR (red squares) vs. the fraction of eigenstate index $ n/L $ for $ h = 3.0 $, $ a = 1.5$ (long-range limit), $ \alpha = 0.2$, and system size $ L = 6765 $. (b) A zoomed-in view of NPR (red squares) and localization length, $ \xi $ (green triangles) versus $n/L$ highlighting the NHCE. (c) Energy spectrum in the complex plane under PBC for the same parameter set. (d) Phase diagram in the $ \alpha $ - $ h $ plane for $ a = 1.5 $, where I, C, D, and L denote the intermediate (blue), comb (white), extended (green), and localized (red) regions, respectively (see text for details) with system size $L=610$. The color-bar is appropriately set for better visibility.}
\label{fig:fig2}
\end{figure}

Now we will focus on to understand this  anomalous spectral features in detail. For this purpose we first plot the IPR (blue circles) and NPR (red squares) as a function of the eigenstate indices ($n/L$) of all the states for $h=3.0$ and $a=1.5$ in Fig.~\ref{fig:fig2}(a). It can be seen that the states in the lower part of the energy spectrum are localized which can be seen as possessing finite values of IPR and vanishing NPR. However, from indices  $n/L=2187/6765$ to $n/L=6114/6765$, the states with finite IPR and states with finite NPR appear intermixed with each other. To obtain better clarity on this atypical distribution of the states we zoom in a region of the spectrum in Fig.~\ref{fig:fig2}(b). It is evident from the figure that all the extended states (finite NPR) are separated from each other by a bunch of localized states (vanishing NPR and finite IPR). To further quantify this feature we plot the localization length (green triangles) for all the states in Fig.~\ref{fig:fig2}(b) which is defined as $\xi=\sqrt{\sum_{j}^{L}(j - j_c)|\psi_j|^2}$, where $|\psi_j|^2$ is the single particle probability density at site $j$ and $j_c=\sum_{j=1}^{L}j|\psi_j|^2$ is the localization center. The large (vanishing) values of $\xi$ confirms this anomalous distribution of the extended (localized) states.  This intermittent occurrence of the extended states in this regime of the  spectrum follows an almost regular pattern and is reminiscent of a comb like feature in the energy spectrum - a pattern which was not seen in any of the quasiperiodic lattice models. We call this feature the non-Hermitian comb effect (NHCE). It is important to note that the entire spectrum does not exhibit NHCE rather the states which take part in the NHCE are separated from the localized states (finite IPR and vanishing NPR) at the lower part ($ < n/L=2187/6765$) and at the upper part ($> n/L=6114/6765$) of the energy spectrum. As a result of this we obtain two MEs which separate the states participating in the NHCE and the localized states and we call them the comb MEs. 

To further understand the spectral behavior of the intermediate phase due to the NHCE we compare the real and imaginary energy eigenvalues of the states. It is well known that in non-Hermitian systems, the energy spectrum is typically complex, in contrast to Hermitian ones. 
Recent studies have discovered `mobility rings'~\cite{Li_mobrings_2024}, which are unique to non-Hermitian systems, adding new complexities to the interplay between spectral structure and localization effects. These intricate MEs form loop-like structures in the complex plane and help clearly distinguish localized states from extended ones within the spectrum. In this context we examine the energy spectrum in the complex plane under PBC at the same parameter values as in Fig.~\ref{fig:fig2}(b) that yields this comb-like structure. 

A highly unusual pattern appears when we compare the real and imaginary energy eigenvalues along with the NPR of the corresponding states as shown in Fig.~\ref{fig:fig2}(c). The localized states, characterized by complex energies and NPR values close to zero (blue dots), construct loop-like structures. These loops encircle the extended states with real energies and finite NPR values (red dots). Rather than forming a distinct boundary between localized and extended states, as commonly observed in other non-Hermitian systems, we find that the extended states lie within the loop formed by the localized states. This behavior reveals a non-trivial character of the spectrum which is in complete contrast with the traditional concept of a sharp MEs in the spectrum, whether real or complex.

The above inference can be summarized in the form of a phase diagram in the $\alpha-h$ plane as shown in Fig.~\ref{fig:fig2}(d) which is obtained by comparing the NPR and IPR profiles of the states (see Sec. S.1). The phase diagram depicts the extended (D), localized (L) and intermediate (I) phases as well as the intermediate phase with NHCE (C) which are denoted by green, red, blue, and white colors, respectively. It is evident from the phase diagram that the NHCE is also seen for larger values of $\alpha$ for which the smaller $h$ is sufficient to establish such effect. 

\begin{figure}[htp]
\centering
\includegraphics[width=1.0\columnwidth]{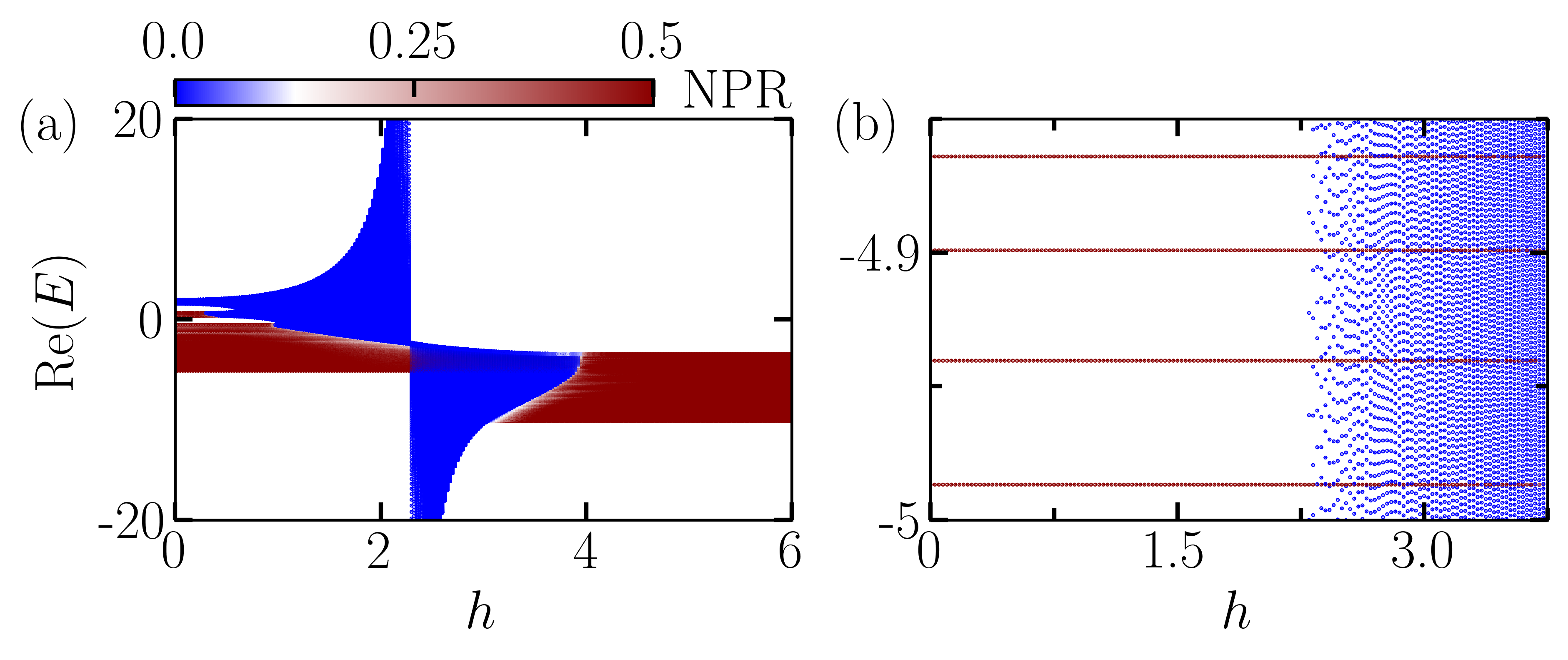}
\caption{(a) Real part of the energy as a function of $ h $ with corresponding NPR values of the states for $ a = 1.5 $ (long-range limit), $ \alpha = 0.2$, and system size $ L = 6765 $. The color bar indicates the NPR values. (b) A zoomed-in view of the energy spectrum, illustrating the "comb effect'', where red-colored energy values correspond to robust extended states passing through the pool of localized states (blue dots). The color-bar is appropriately set for better visibility.}
\label{fig:fig4}
\end{figure}

The origin of the NHCE obtained here can be understood by carefully analyzing the spectral behavior of the system. For this purpose we plot the energy eigenvalues of all the states as a function of $h$ for $\alpha=0.2$ and $a=1.5$ in Fig.~\ref{fig:fig4}(a) along with the NPR of the corresponding states (denoted in colors). It can be seen that for up to $h\approx 3.9$, the spectrum exhibits well separated extended (red) and localized (blue) states at the lower and upper part of the spectrum, respectively separated by the ME. However, when we vary the complex phase $h$ with a fixed disorder strength, the potential undergoes a sharp transition at the critical point $ h=h_c=\text{ln} |(1+\sqrt{1-\alpha^2})/\alpha|$ (for $\alpha=0.2$, $h_c\approx2.3$). Near this point, the potential abruptly inverts, rapidly shifting from large positive to large negative values. This sudden change creates a discontinuity that significantly affects the distribution of eigenenergies. As shown in Fig.~\ref{fig:fig4}(a), this inversion in the potential induces a similar inversion in the eigenenergy spectrum. As a result of this an intermixed pattern of localized (blue dots) and extended (red dots) states appear for a region of $h$ between $2.3 \lesssim h \lesssim 3.9$, which is responsible for the NHCE. To better understand this we enlarge the intermixed region of the spectrum in Fig.~\ref{fig:fig4}(b). It can be seen that robust extended states (red dots) with consistent energies persist in this range as they pass through regions of localized states (blue dots). As a result a layered structure of extended states amid clusters of localized states appear which bring in the features of the NHCE.

It is now important to note that for the NHCE to occur two key features have to be satisfied; (i)  the ME should reach the point $h=h_c$ i.e., some extended states must exist before \( h \) reaches $ h_c $. (ii) These extended states maintain robust energy values (red) even as they intersect localized regions (blue), a feature unique to NHCE. This stability in the energy of extended states, as they move through localized regions, reinforces the comb effect, producing a layered structure where extended states recur amid clusters of localized states. 

If all states localize before reaching $ h_c $, the comb effect cannot occur. This explains the absence of NHCE in the $ a = 10 $ case for $ \alpha = 0.2 $. While this strong potential is sufficient to fully localize the spectrum in the short-range hopping case (see Fig.~\ref{fig:fig1}(a)), the inclusion of long-range hopping promotes the existence of extended states up to the critical value $h = h_c$ (see Fig.~\ref{fig:fig1}(b) and (c)), thereby supporting the NHCE. In the case of the short-range hopping limit, for smaller values of $\alpha$, the critical value $h_c$ is relatively high. Consequently, all the extended states become localized before the mobility edge reaches $h_c$, preventing the observation of NHCE in the short-range limit for smaller values of $\alpha$ (as shown in Fig.~\ref{fig:fig1}(a)). However, for larger values of $\alpha$, the $h_c$ value shifts consistently toward smaller $h$, allowing the mobility edge to reach $h_c$. This shift of $h_c$ facilitates the observation of NHCE even in the short-range hopping regime for larger $\alpha$ values.


\begin{figure}[t]
\centering
\includegraphics[width=1.0\columnwidth]{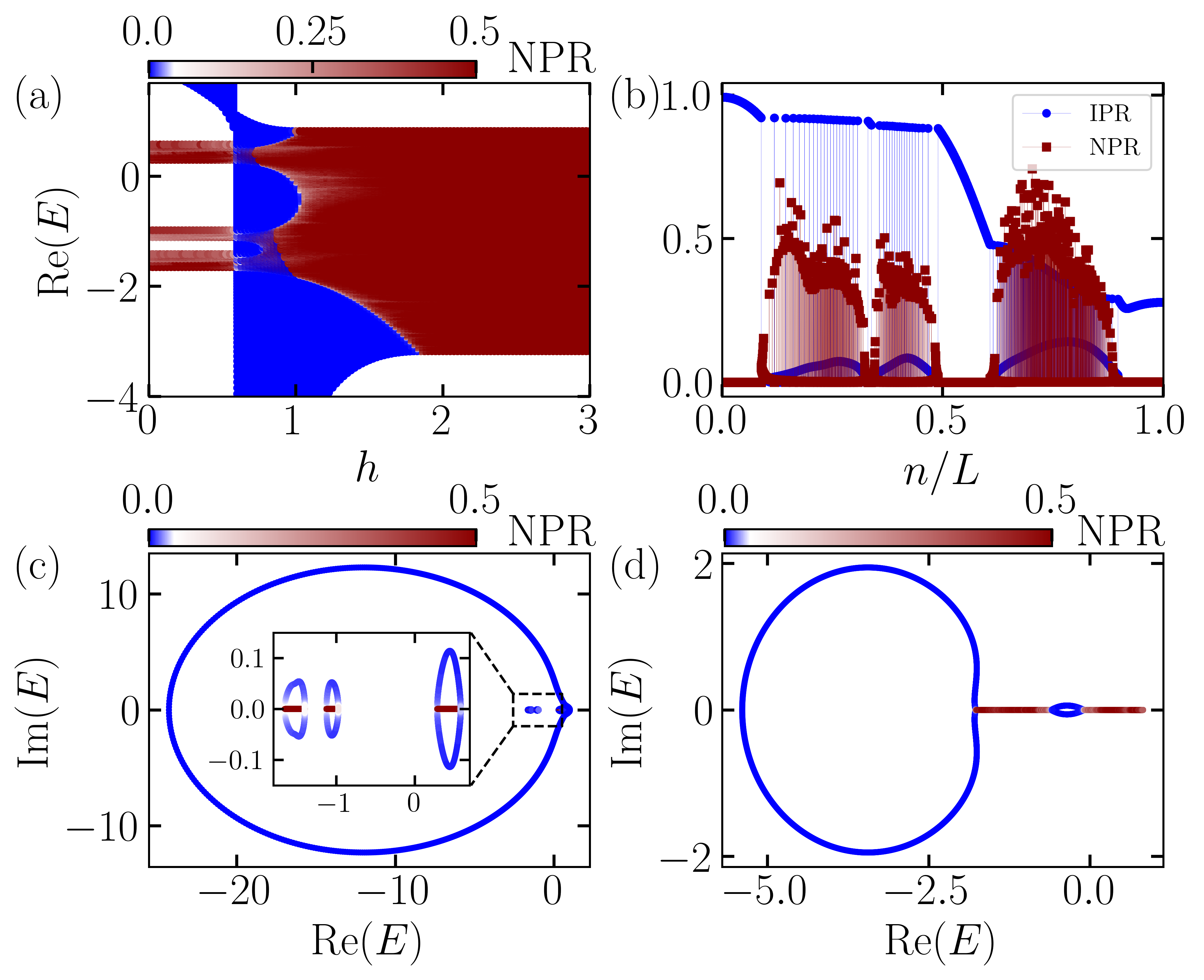}
\caption{(a) Real part of the energy vs. $ h $ with their corresponding NPR values (shown as colors) for  $ a = 10$ (short-range limit), $ \alpha = 0.85 $, and $ L = 6765 $. (b) IPR (blue circles) and NPR (red squares) vs. fraction of eigenstate index $ n/L $ at $ h = 0.675$ depicting more complex  NHCE as compared to the long-range scenario (compare Fig.~\ref{fig:fig2}). (c), (d) Show the Energy spectrum in the complex plane under PBC with corresponding NPR of the states for $ h = 0.675 $ and $ h = 1.0 $, respectively. The inset in (c) shows the zoomed view of the complex eigenspectra. The color-bar is appropriately set for better visibility.}
\label{fig:fig3}
\end{figure}

To visualize this we plot the real energies as a function of $h$ along with the NPR (color coded) of the states for $a=10$ and $\alpha=0.85$ while keeping $\lambda=1$ as before in Fig.~\ref{fig:fig3}(a). This depicts multiple bands of extended states reach up to the critical $h_c\approx0.585$ and remain robust after that which is the condition for the spectrum to exhibit NHCE. To see the signature of the NHCE we compare the NPR (red squares) and IPR (blue circles) of all the states in Fig.~\ref{fig:fig3}(b). The oscillatory pattern in the NPR and IPR confirms the emergence of multiple regions in the spectrum exhibiting the NHCE. To gain more insight on such anomalous features, we plot the eigenvalue spectrum in the complex plane along with the corresponding NPR values in Fig.~\ref{fig:fig3}(c). This reveals three small loops within a larger loop, all characterized by NPR values near zero, corresponding to the localized states (blue dots). These smaller loops originate from the oscillating IPR values of localized states within the three comb regions in the spectrum. Inside each of these smaller loops, we observe some extended states with real energies and finite NPR values (red dots). Note here that this unusual feature is not only more richer compared to the long-range case but also completely different from the situation with well-defined MEs. For comparison, we plot the energy spectrum slightly outside the comb region at $ h = 1.0 $, where we see well-defined MEs as shown in Fig.~\ref{fig:fig3}(d). Here, the localized states (blue) form loop-like structures, which are separated from the extended states (red dots), leaving outside the loops with real energy values.

\paragraph*{Conclusion.-}
\label{sec:Conclusion}

In this study, we have examined the unique properties of a non-Hermitian quasiperiodic chain with both long- and short-range hopping, uncovering an atypical intermediate phase where the extended and localized states appear intermixed with each other. In such an intermediate phase the extended states in pairs are found to be separated from each other by a cluster of localized states - a feature contrasting to the traditional expectation of a sharp boundary between localized and extended phases. We call this almost regular pattern in the spectrum the non-Hermitian comb effect (NHCE). 
This non-trivial spectral feature provides new insights into the unconventional behaviors of intermediate phases and eigenspectra in non-Hermitian systems. 
It also goes beyond the conventional understanding of the spectral behavior of the quasiperiodic systems, which may open up possibilities for further exploration in this direction. 
It will be informative to uncover the existence of such NHCE in other lattices, explore their spectral statistics, conduct multifractal analysis, and examine the implications for dynamical behavior.
Furthermore, an exploration of the spectral topology of the energy spectrum in the complex plane in the comb region holds the potential to uncover novel topological phenomena, enriching our understanding of non-Hermitian systems. Also, investigating the robustness of the NHCE in the presence of interactions could provide deeper insights into its many-body effects. 
Most importantly, with recent advances in accessing quasiperiodic lattices with complex phases in systems like electric circuits~\cite{Zeng_GMEs_NHGAA_expt_2020}, photonic lattices~\cite{Tong_liu_2020, Longhi_prl_2019, ghatak2024} and periodically driven system~\cite{Weidemann2022}, our findings can be observed in experimental set ups. 

\paragraph*{Acknowledgement.-}
T.M. acknowledges support from Science and Engineering Research Board (SERB), Govt. of India, through project No. MTR/2022/000382 and STR/2022/000023.

\bibliography{ref}

\newpage

\begin{center}
\section*{ \bf Supplemental Material}
\end{center}

\renewcommand{\thesubsection}{S}
\renewcommand{\theequation}{S-\arabic{equation}}
\setcounter{equation}{0}

In this supplementary material, we characterize different phases shown in main text. We also show the phase diagram for further longer-range hopping.


\begin{center}
{\it {\bf S.1. Phase diagram for 
 $a=1.5$ }}\par
 \label{sec:S1}
\end{center}


\begin{figure}[h]
\centering
\includegraphics[width=1.0\columnwidth]{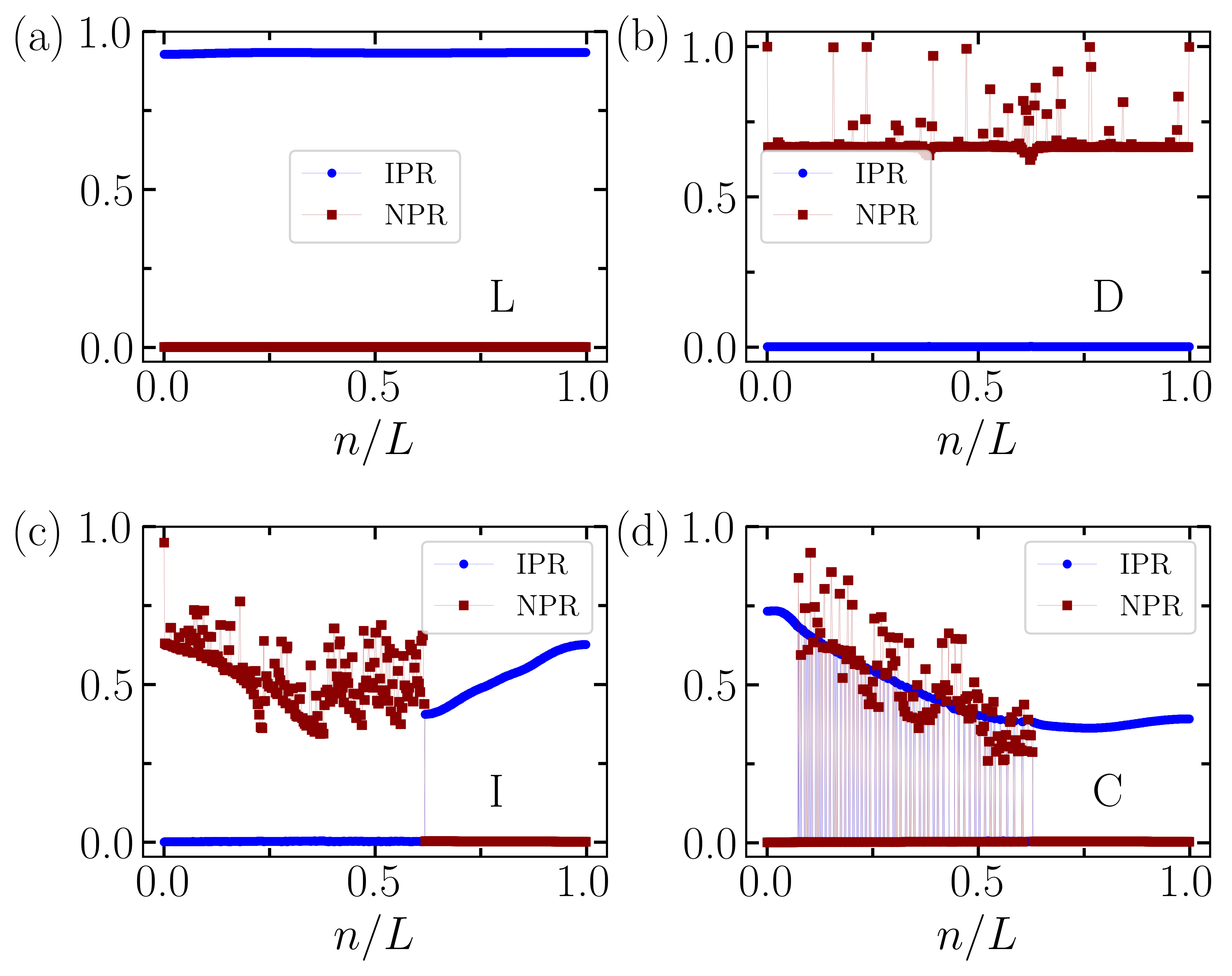}
\caption{IPR and NPR plotted as a function of eigenstate index. For (a) localized, (b) extended, (c) intermediate and (d) comb with $(\alpha, h)$ as $(0.05, 5.0)$, $(0.8, 5.0)$, $(0.2, 0.25)$ and $(0.8, 1.0)$, respectively.}
\label{fig:fig6}
\end{figure}

The phases in the phase diagram shown in Fig.~\ref{fig:fig2} of the main text highlights the  extended (D, green), localized (L, red), intermediate (I, blue), and comb (C, white) phases for $a=1.5$ and $\lambda=1$. These phases are identified by examining the inverse participation ratio (IPR) and normalized participation ratio (NPR) of the entire spectrum at each $\alpha, h$ value. The phases are characterized by comparing the IPR and NPR, which vary distinctly across different phases. In the localized phase, the IPR remains finite, and the NPR approaches zero for all eigenstates, as evident in Fig.~\ref{fig:fig6}(a) ($\alpha=0.05$, $h=5.0$), reflecting clear localization. In the extended phase, the IPR vanishes while the NPR remains finite throughout the spectrum, as shown in Fig.~\ref{fig:fig6}(b) ($\alpha=0.8$, $h=5.0$), indicating extended states. The intermediate phase exhibits a sharp boundary between localized and extended states, with part of the spectrum showing a finite IPR and the rest approaching zero, complemented by opposing NPR behavior, as demonstrated in Fig.~\ref{fig:fig6}(c) ($\alpha=0.2$, $h=0.25$). Finally, the comb phase reflects the intermixing of localized and extended states through oscillatory behavior in both IPR and NPR, as illustrated in Fig.~\ref{fig:fig6}(a) ($\alpha=0.8$, $h=1.0$).

\begin{center}
{\it {\bf S.2. Phase diagram for 
 $a=0.5$}}\par
\end{center}

\begin{figure}[h]
\centering
\includegraphics[width=0.75\columnwidth]{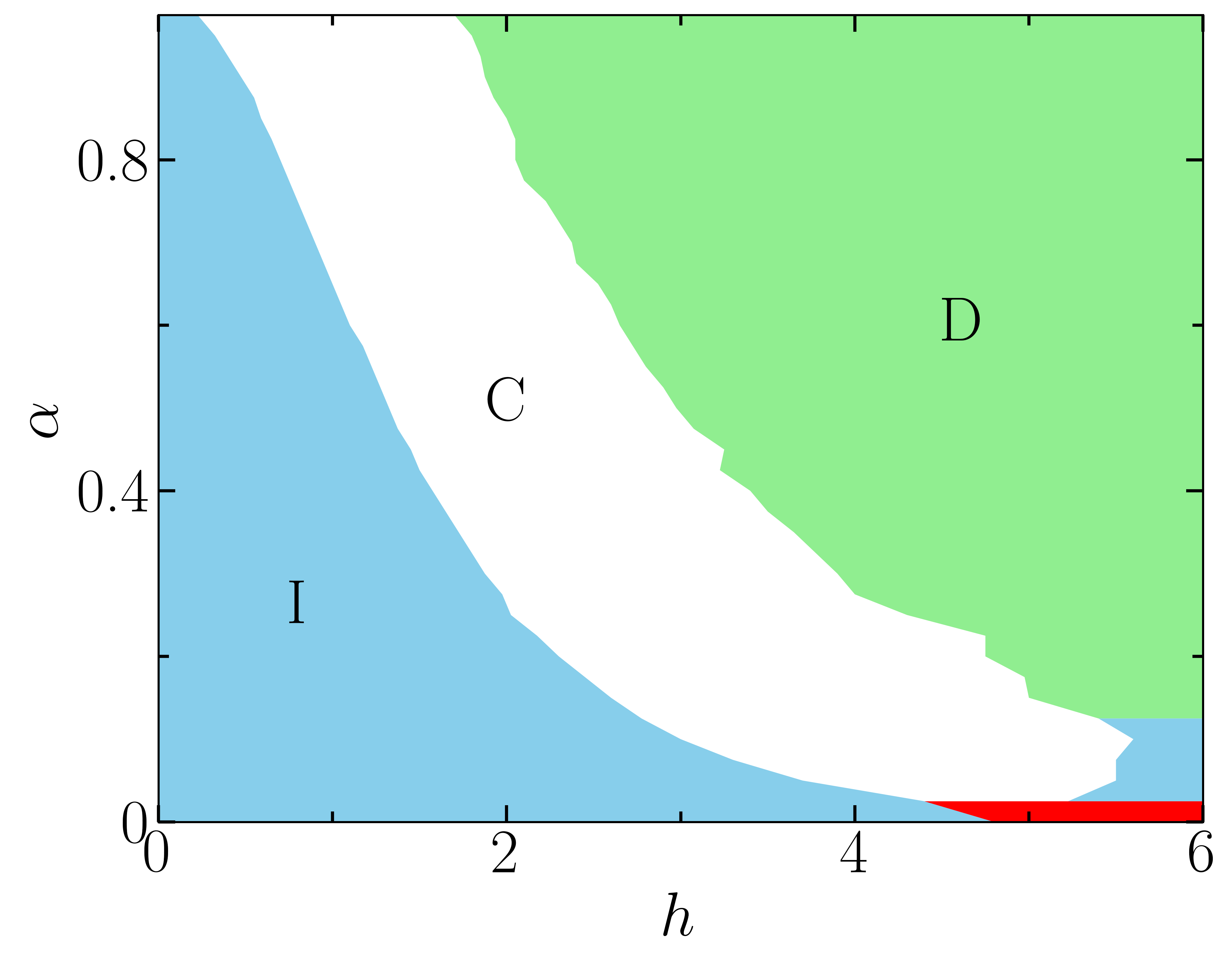}
\caption{Phase diagram in $\alpha-h$ plane showing the localized (L), extended (D), intermediate (I) and comb (C) represented by red, green, blue and white regions, respectively. Here, the system parameters are $a=0.5$, $\lambda=1$ and system size $L=610$.}
\label{fig:fig7}
\end{figure}

To explore the effect of further longer-range hopping we plot the phase diagram for the power-law hopping index $a=0.5$ as shown in Fig.~\ref{fig:fig7}. Similar to the phase diagram shown in the main text, here also we obtain four distinct phases: extended (D, green), localized (L, red), intermediate (I, blue), and comb (C, white). However, compared to case of $a=1.5$, in this case we obtain an enlarged comb phase which even appears at smaller values of $\alpha$. Moreover, the localized region shrinks due to the effect of stronger long-range hopping.

\end{document}